%% file: main.tex
\documentclass[twocolumn,superscriptaddress]{revtex4-1}
\date{\today}
\usepackage{epsfig}
\usepackage{epstopdf}
\usepackage{subfigure}
\usepackage{graphicx}
\usepackage{dcolumn}
\usepackage{bm}
\usepackage[colorlinks,linkcolor=blue,hyperindex,CJKbookmarks]{hyperref}
\usepackage{float}
\usepackage{hyperref}
\usepackage{comment}
\usepackage{xcolor}
\usepackage[textsize=footnotesize,textwidth=0.5in]{todonotes}
\usepackage{soul} 
\makeatletter

\newcommand{\kiro}   {K$_2$IrO$_3$}
\newcommand{\kiroxy}   {K$_x$Ir$_y$O$_2$}
\newcommand{\niro}   {Na$_2$IrO$_3$}
\newcommand{\aliro}   {$\alpha$-Li$_2$IrO$_3$}
\newcommand{\arucl}   {$\alpha$-RuCl$_3$}

\begin{document}

\title{Large off diagonal exchange couplings and spin liquid states in
$\mathbf{C_3}$ symmetric iridates}

\author{Ravi Yadav}
 \affiliation{IFW Dresden, Helmholtzstr. 20, 01069 Dresden, Germany}

\author{Satoshi Nishimoto}
 \affiliation{IFW Dresden, Helmholtzstr. 20, 01069 Dresden, Germany}
 \affiliation{Department of Physics, TU Dresden, D-01062, Dresden, Germany}

\author{Manuel Richter}
 \affiliation{IFW Dresden, Helmholtzstr. 20, 01069 Dresden, Germany}
 \affiliation{Dresden Center for Computational Materials Science (DCMS), TU Dresden, 01062 Dresden, Germany}

\author{Jeroen van den Brink}
 \affiliation{IFW Dresden, Helmholtzstr. 20, 01069 Dresden, Germany}
 \affiliation{Department of Physics, TU Dresden, D-01062, Dresden, Germany}

\author{Rajyavardhan Ray}
 \email{r.ray@ifw-dresden.de}
 \affiliation{IFW Dresden, Helmholtzstr. 20, 01069 Dresden, Germany}
 \affiliation{Dresden Center for Computational Materials Science (DCMS), TU Dresden, 01062 Dresden, Germany}

\date{\today}

\begin{abstract}
    Iridate oxides on a honeycomb lattice are considered promising candidates for
    realization of quantum spin liquid states. We investigate the
    magnetic couplings in a structural model for a honeycomb iridate
    \kiro, with $C_3$ point group symmetry at the Ir sites, which
    is an end member of the recently synthesized iridate family
    {\kiroxy}. 
    Using \textit{ab-initio} quantum chemical methods, we elucidate the subtle
    relationship between the real space symmetry and magnetic anisotropy
    and show that the higher
    point group symmetry leads to high frustration with strong magnetic anisotropy driven
    by the unusually large off-diagonal
    exchange couplings ($\Gamma$'s) as opposed to other spin-liquid candidates considered 
    so far. Consequently, large quantum fluctuations imply lack
    of magnetic ordering consistent with the experiments.
    Exact diagonalization calculations for the fully anisotropic
    $K$-$J$-$\Gamma$ Hamiltonian reveal the importance
    of the off-diagonal anisotropic exchange couplings
    in stabilizing a spin liquid state and highlight an alternative route
    to stabilize spin liquid states for ferromagnetic $K$. 
\end{abstract}

\maketitle

\section{Introduction}
The possibility to realize a spin liquid (SL) state in condensed matter
systems is being intensively investigated from both theoretical and
experimental standpoints \cite{Banerjee16,Helgaker2000,Kim14}. SL states are characterized by large
degeneracy in the ground state, suppression of
long-range (magnetic) order, and cannot be described by the broken
symmetries associated with conventional magnetic ground states \cite{Ir214_lupascu_14}. 
In this regard, most promising candidates are the spin-orbit-driven Mott
insulators on a honeycomb lattice \cite{chaloupka10}, such as
{\niro} \cite{choi12,chun15}, {\aliro} \cite{singh12a,Mazin2013}, and
{\arucl} \cite{plumb14,banerjee2016,Yadav16}. Due to
$d^5$ configuration in the presence of octahedral crystal field
environment and spin-orbit coupling (SOC),
the low-energy electronic properties is typically described by an effective
nearest-neighbor Kitaev-Heisenberg Hamiltonian ($K$-$J$ Hamiltonian) 
acting in the $j_{\rm eff} = 1/2$ subspace. 
Large ratio of Kitaev to Heisenberg couplings, $K/J \gg 1$, found in these
materials imply strong magnetic anisotropy and proximity to a SL
state. However, presence of extended range magnetic couplings and
{small but finite} $J$ renders a long-range magnetic order
instead \cite{Kimchi11,rau14,Katukuri14,Nishimoto16}.

Most of the works in these honeycomb lattice iridates have, therefore, focused on
studying the effects of external stimulus, such as magnetic field
\cite{Yadav16,banerjee2018}, hydrostatic pressure \cite{Gael18,Majumder18,Simutis2018}, 
trigonal distortions \cite{Nishimoto16}, as well as chemical substitutions and
doping \cite{Yadav2018_PRL,koitzsch2017}, in order to tune the magnetic
couplings favorably for a SL state. 
Very recently, a new family of honeycomb lattice iridates,
{\kiroxy} has been
synthesized and in a substantial range of concentrations, including
the end member {\kiro}, a structural model featuring a $C_3$ point group (PG)
symmetry at the Ir sites was proposed \cite{kiro_expt18,kiro_expt18_mag}. 
The magnetic susceptibility measurements suggest that no
long-range order or spin freezing develops down to 1.8$K$ while the specific heat is finite at
low temperatures, implying the possibility of a gapless quantum SL
state.\cite{kiro_expt18_mag}

Here, we report the influence of $C_3$ point group
symmetry at the transition metal ion site on 
 the nearest neighbor (NN) magnetic 
 interactions between Ir atoms. Starting with the proposed
 structural model for {\kiro}, we discuss the evolution of the
 magnetic couplings with deviations from the high symmetry structure and implications for 
the stability of a SL state. The NN magnetic couplings are obtained using quantum chemistry electronic 
structure calculations, performed for a crystal structure with optimized atomic
positions as obtained within density functional theory (DFT). We find
unusually large off-diagonal exchange couplings: $\Gamma_{xy} \sim 5$ meV and
$\Gamma_{yz} \sim -9$ meV (in the local Kitaev frame),
approximately 10 times larger than in {\niro}. 
At the same time, 
while the $K$ and $J$ exchange terms are smaller, $K/J$ ratio
is comparable to other honeycomb
iridates, thus, motivating a $K$-$J$-$\Gamma$ Hamiltonian as an
appropriate model to capture the underlying physics in {\kiro}. 
We identify the origin of such large $\Gamma$'s as
constraints on the relative orientation of the 
O-O pairs within the ${\rm IrO_6}$ octahedra due to $C_3$ PG symmetry at the Ir sites
imposed by the large K ions. The large $\Gamma$-driven magnetic anisotropy implies
strong magnetic frustration and is responsible for suppression of magnetic
ordering, observed experimentally.

We note that there is an emerging consensus regarding the importance
    of $\Gamma$ terms in understanding the
    magnetic interactions in honeycomb lattice Mott
    insulators. Therefore, models involving such terms are being
    investigated. For example, an anisotropic $\Gamma$ as well as a $K$-$\Gamma$ model on a honeycomb lattice
supports a SL state \cite{gamma_SL2017,Catuneanu2018}.
The $\Gamma$-terms are also found to be important in explaining the recent
neutron scattering and high-temperature magnetic susceptibility
experiments in {\arucl} \cite{Lampen-Kelly2018,Gohlke2018}.
However, these works consider only one component of $\Gamma$ per bond,
as opposed to our findings for the $C_3$-symmetric
structure.

Our exact diagonalization (ED) calculations using the quantum chemical NN
magnetic couplings for {\kiro} reveal that the fate of a SL state in
the fully anisotropic 
$K$-$J$-$\Gamma$ Hamiltonian is
determined by the relative signs of $\Gamma_{ij}$'s: 
a $\Gamma$-driven SL state is found to be stable for small values of
$\Gamma_{ij} < 0$ and spread over a large region in the
$\Gamma_{xy}$-$\Gamma_{yz}$ plane.
Moreover, the ferromagnetic$-$Kitaev-SL$-$stripy
path of the $K$-$J$ model \cite{chaloupka10} is also recovered at finite $\Gamma_{ij}$'s.
Further inclusion of extended range
Heisenberg couplings suggest competing magnetic orders in {\kiro} which may have interesting
implications for magnetism. 

\section{Results and Discussions}
\subsection{Structural details}
In the proposed structural model,
{\kiroxy} ($1 \lesssim x/y < 2$) crystallizes in the high symmetry space group $P6_322$ 
(\# 182) \cite{kiro_expt18} as opposed to $C2/m$ for the related
iridate {\niro} \cite{choi12}. For brevity, we focus on the
stoichiometric end
member {\kiro}, corresponding to $x = 4/3$ and $y = 2/3$.
Starting from the experimental values of the external parameters, the
atomic positions were optimized using DFT utilizing the space group
symmetries (see Methods and Supplemental Material). 

\begin{figure}
    	\centering
    	\includegraphics[angle=0,width=0.235\textwidth]{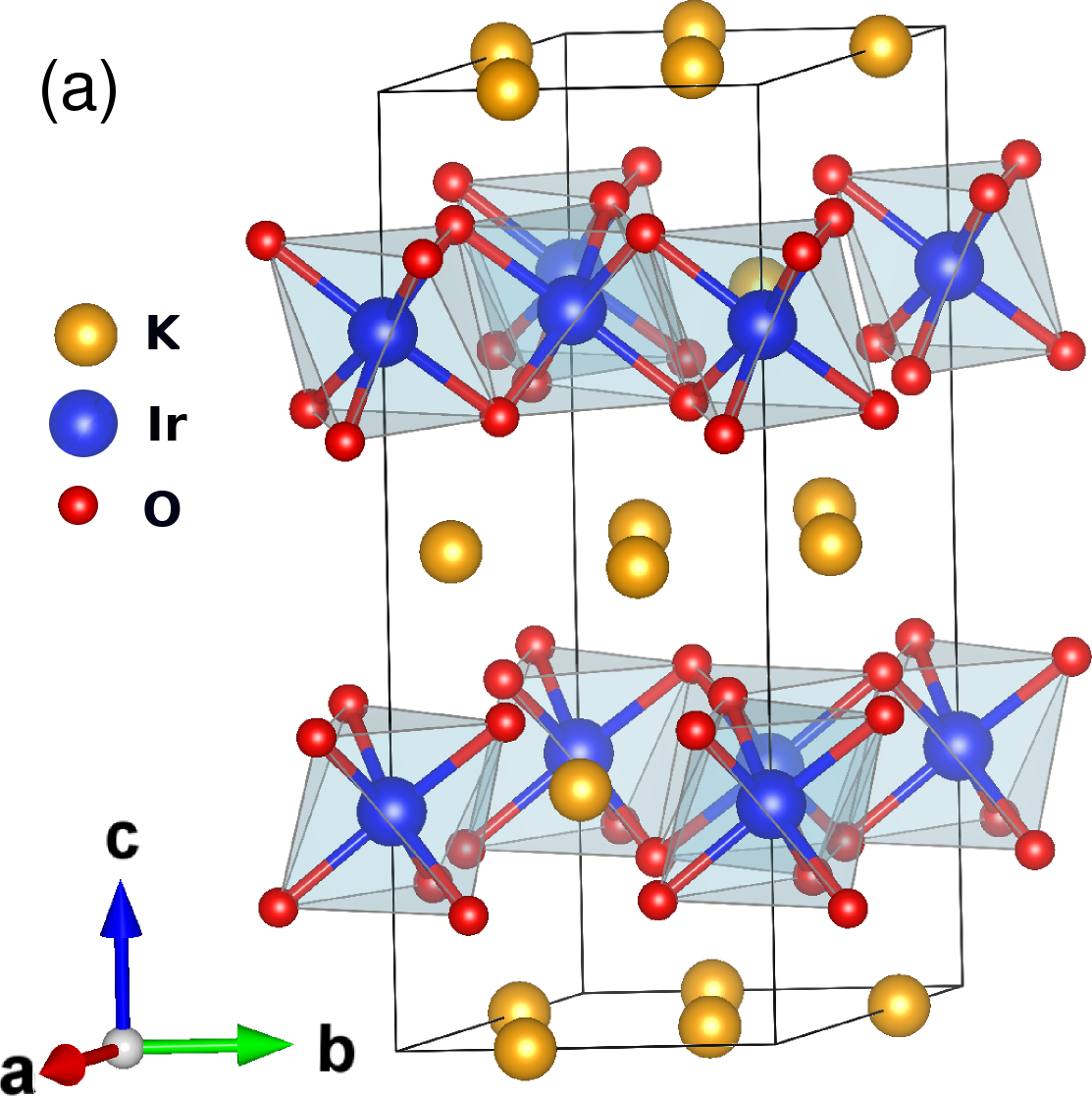}
        \hspace{-0.2cm}
    	\includegraphics[angle=0,width=0.245\textwidth]{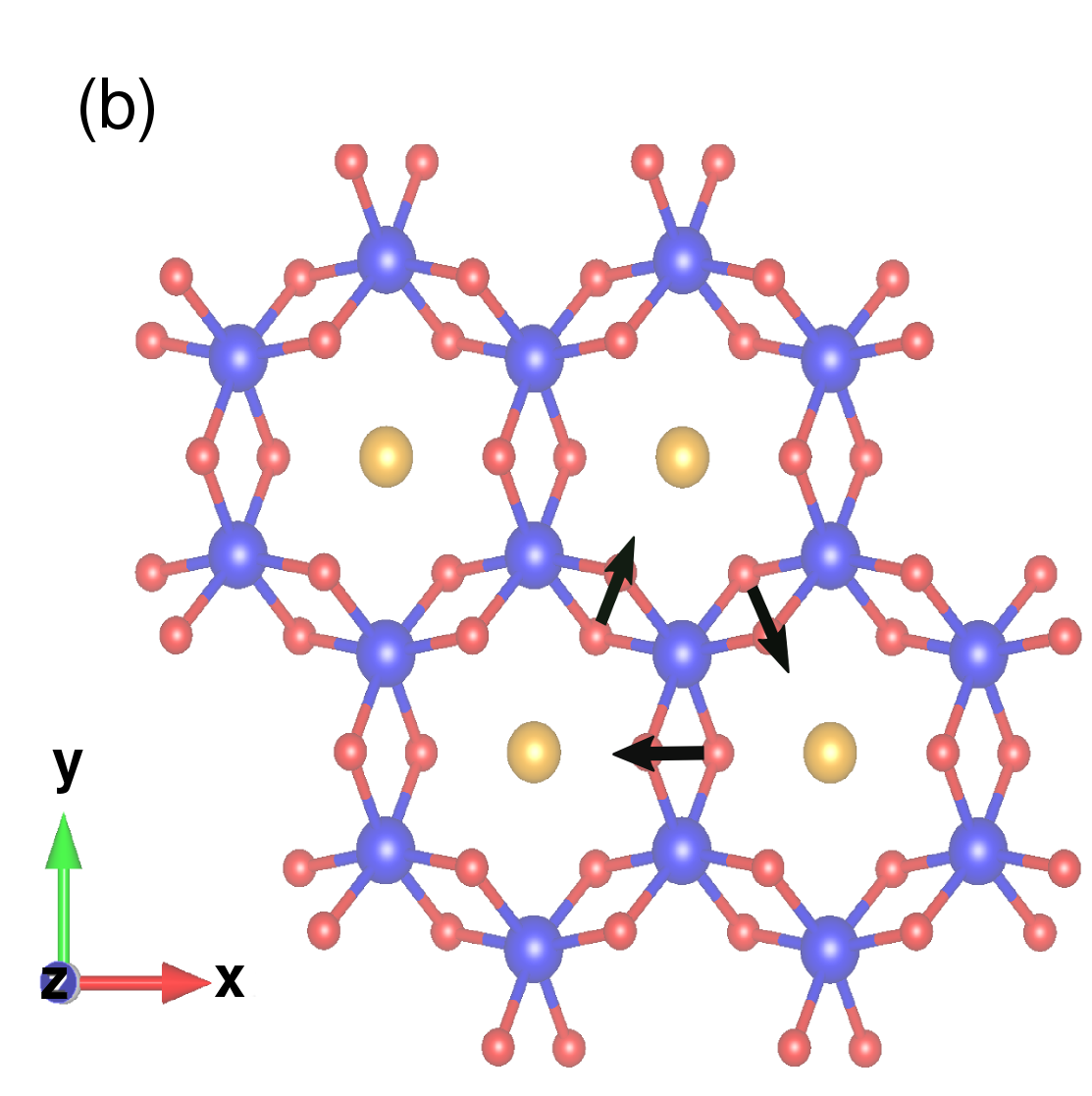}
        \caption{\small
        Crystal structure
            of {\kiro}: (a) shows the alternating layers formed of K ions, and
            hexagonal network formed by ${\rm IrO_6}$ octahedra with K ions at the center of the
            hexagons. 
            (b) Top view of the honeycomb plane spanned by the ${\rm IrO_6}$
            octahedra and K ions at center. Manifestation of the $C_3$ PG
            symmetry: the in-plane projections of the O-O links in each Ir$_2$O$_2$
            plaquette are oriented $120^{\circ}$ with respect to each
            other (shown by arrow from the O atoms above the Ir plane to O atoms below the Ir plane). 
        }
    	\label{fig:str_kiro}
\end{figure}

The resulting structure features alternating layers of non-magnetic cations 
(K ions, in this case) and a two-dimensional/planar honeycomb network of ${\rm IrO_6}$
octahedra. Within the planar honeycomb network, the transition metal
(TM) ions span
a regular honeycomb lattice and share O-O edges between them (see Fig.
\ref{fig:str_kiro}). Within each honeycomb plane, one K ion is also present
at the center of the hexagons. As compared to {\niro}, the
TM-ligand distance $d_{\rm Ir-O} = 2.07$ {\AA} is comparable, while
the TM-TM distance $d_{\rm Ir-Ir}= 3.05$ {\AA} and the
TM-O-TM angle $\angle {\rm Ir-O-Ir} = 95.06^{\circ}$ are much smaller
(see the Supplemental Material). 
Further notable differences with the
related iridate {\niro} are (i) the larger inter-layer separation: the two
Ir (or, K) layers are separated by approximately 6.8 {\AA}, as opposed to 5.6 {\AA} in {\niro};
(ii) Each IrO$_6$ octahedron has a $C_3$ symmetry about the `c' axis centered
at the Ir atoms. As a result, a highly symmetric
real space structure is realized where the projection of O-O links at
every Ir-Ir bonds on the honeycomb plane are at $120^{\circ}$ relative to each other, shown in
Fig. \ref{fig:str_kiro}(b).

\subsection{Electronic properties and magnetic couplings}
The Ir-$5d$ levels in the presence of an octahedral ligand field split
into $e_{\rm g}$ and $t_{\rm 2g}$ levels, with the latter lying at
significantly lower energy \cite{Gretarsson13} (also see
the Supplemental Material). 
The large crystal field splitting results in $t_{\rm 2g}^5$ configuration
yielding effectively one hole per site.
Strong spin-orbit coupling further splits the $t_{\rm 2g}$ states into 
fully occupied $j_{\rm eff} =$ $3/2$ and half-filled $j_{\rm eff} =$ $1/2$ 
states \cite{Jackeli09, Abragam70, Kim08}.
Deviations from an ideal octahedral environment may lead to some admixture
between these $j_{\rm eff}$ states. 

The nearest-neighbor (NN) magnetic interactions were derived using
quantum chemistry calculations performed on embedded clusters consisting
of two edge-shared octahedra (Ir$_2$O$_{10}$ units) plus their NN octahedral
units and NN K ions (see Methods).
At each Ir-Ir bond in {\kiro}, 
this structural unit has an approximate $C_{2h}$ point-group
symmetry. Since the deviation from the $C_{2h}$ symmetry is very small
($\lesssim 0.1$ \%; see the Supplemental Material), an antisymmetric Dzyaloshinskii-Moriya (DM)
term, which is not allowed in $C_{2h}$ symmetry due to the presence of
an inversion center, is expected to be negligible. Therefore, a generalized
bilinear Hamiltonian between a pair of pseudospins 
$\mathbf{\tilde S}_i$ and $\mathbf{\tilde S}_j$ is obtained for the
$C_{2h}$ symmetry of the structural unit \cite{Katukuri14}:
\begin{equation}
    {\cal H}^{(\gamma)}_{ij} =J\, \mathbf{\tilde{S}_i} \cdot \mathbf{\tilde{S}_j}
           +K \tilde{S}^\gamma_i \tilde{S}^\gamma_j
           +\sum_{\alpha \neq \beta} \Gamma_{\!\alpha\beta}(\tilde{S}^\alpha_i\tilde{S}^\beta_j +
                                                          \tilde{S}^\beta_i \tilde{S}^\alpha_j), \ \
\label{eqn:bilinear_qc}
\end{equation} 
where, $J$ and $K$ are the Heisenberg and Kitaev exchange couplings,
respectively, and $\Gamma_{\alpha\beta}$  coefficients are the
off-diagonal components of the symmetric anisotropic exchange matrix,
with $\alpha,\beta = {x,y,z}$. 
A local Kitaev reference frame is used, such that for each Ir-Ir
bond, the $z$-coordinate is perpendicular to the Ir$_2$O$_2$ plaquette.

\begin{table}
    \centering
    \caption{\small
    Nearest neighbor magnetic couplings (in meV) for {\kiro}
    and as obtained from spin-orbit MRCI calculations for 
    the DFT-optimized structure. For comparison, the corresponding
    values for bonds {\it B1} (top) and {\it B2} (bottom) in {\niro} are also mentioned (taken
from Ref. \cite{yadav2018}). 
    }
\label{table:couplings}
\begin{tabular}{lccrrc}
 \hline \hline
     ${\rm A_2IrO_3}$ \hspace{ 0.10cm} &\hspace{ 0.10cm} $\angle $Ir-O-Ir & \hspace{ 0.15cm}  $K$  &
     \hspace{0.15cm} $J$ &\hspace{0.15cm}  $\Gamma_{xy}$   & \hspace{0.10cm} $\Gamma_{yz}$= $-\Gamma_{zx}$ \\
      \hline
      \vspace{-0.1cm}\\
     A = K\hspace{0.15cm}  &\hspace{ 0.15cm} $95.0^\circ(\times3)$ &\hspace{ 0.15cm} $-$6.3   & \hspace{0.10cm} $1.3$  &\hspace{0.15cm} $\textbf{5.2}$  & $\textbf{--8.9}$ \\[0.2cm]
     A = Na\hspace{0.15cm} &\hspace{ 0.15cm} $99.5^\circ(\times1)$ & \hspace{0.15cm} $\textbf{--20.8}$   & \hspace{0.10cm} $\textbf{5.2}$ &\hspace{ 0.15cm}  $-$0.7  &  $-0.8$ \\[-0.02cm]
                           &\hspace{ 0.15cm} $98.0^\circ(\times2)$
      & \hspace{0.15cm} $\textbf{--15.6}$   & \hspace{0.10cm} $\textbf{2.2}$ &\hspace{ 0.15cm}  $-$1.1  & \hspace{0.1cm} $0.8$ \\
        
         \hline
         \hline
 \end{tabular}
\end{table}

The magnetic couplings were determined by
mapping the {\it ab initio} data, obtained in the multi-reference
configuration-interaction (MRCI) calculations including spin-orbit
effects, onto the above effective spin Hamiltonian (Eq. (1)) following the scheme
detailed in Ref. \cite{Bogdanov15,Yadav16,Yadav17} and outlined later.
Note that such a computational procedure has been successfully applied to
other spin-orbit driven Mott insulators 
\cite{Bogdanov2012,Katukuri14,Bogdanov15,Katukuri16,Yadav16,Yadav17,Yadav2018_PRL,Yadav2019_ChemSc}.

The magnetic couplings thus obtained are listed in Table
\ref{table:couplings}. The most striking aspect is the unusually large
off-diagonal exchange couplings, $\Gamma$-terms, 
approximately 10 times as compared to {\niro}. 
This is accompanied by much smaller $K$ and $J$. The ratio
$K/J$, however, is comparable to the corresponding value in {\niro},
implying that {\kiro} is magnetically very frustrated. 
This further suggests that the magnetic anisotropy in {\kiro} is large
and dominated by the $\Gamma$-terms, unlike any other known spin-orbit-driven Mott insulator
on a honeycomb lattice. Consequently, a fully anisotropic $K$-$J$-$\Gamma$ Hamiltonian model would be
necessary to describe this system. It is interesting to note that the magnetic 
couplings listed in Ref. \cite{kiro_expt18} also suggest large
off-diagonal couplings similar to our {\it ab initio} results. The
mismatch in the magnitude of exchange couplings
can be attributed to
different methods employed, as also noticed in previous studies on honeycomb
lattice Mott insulator {\arucl}.\cite{Yadav16,Winter16}

\subsection{Role of $\mathbf{C_3}$ symmetry and inter-layer species}
\begin{figure}
    \centering
    \includegraphics[angle = 0,width=0.40\textwidth]{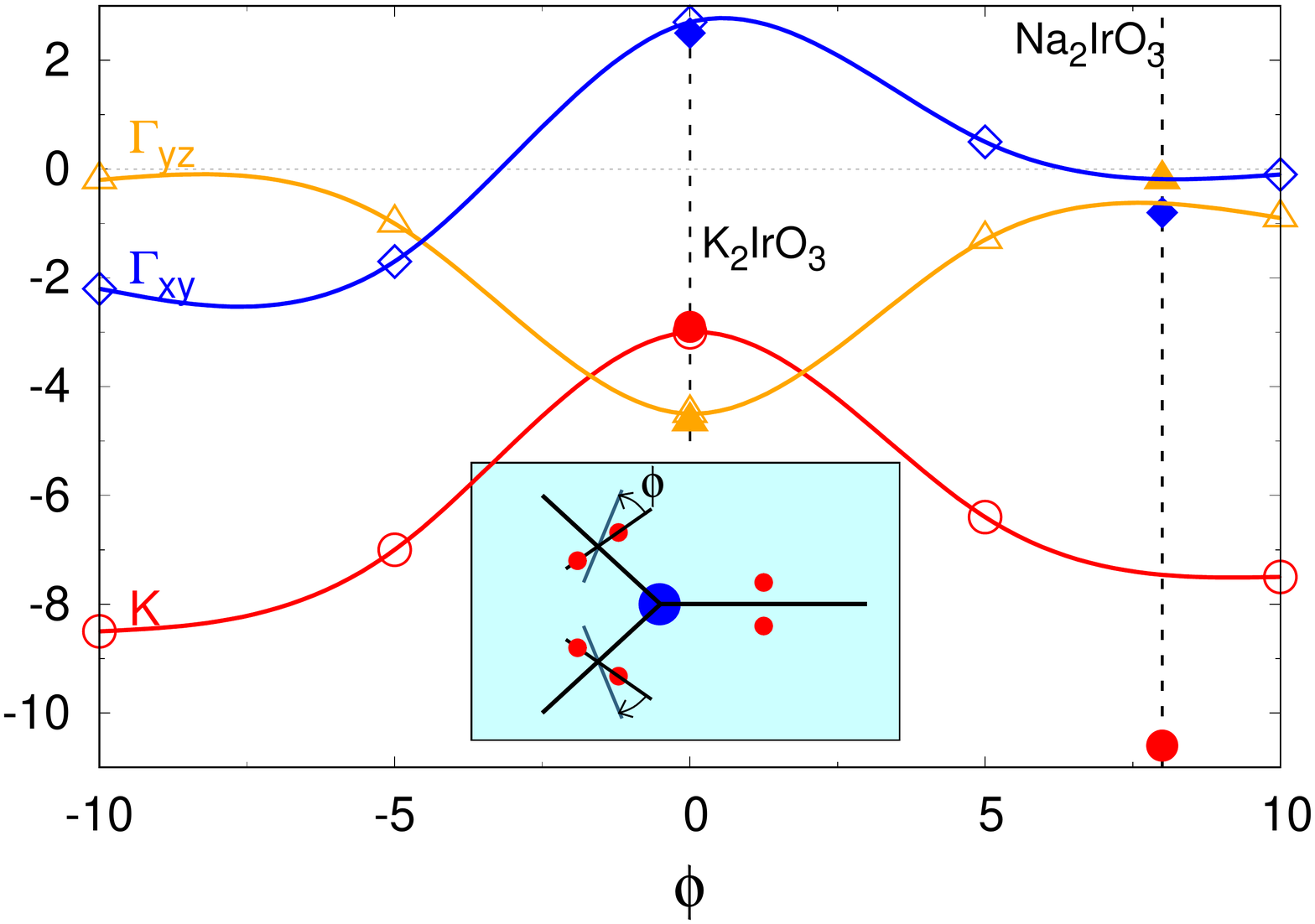}
    \caption{\small
      Effects of $C_3$ point group symmetry on magnetic couplings: Magnetic exchange couplings (in meV) as a function of the
    relative twist angle $\phi$ of the O atoms away from the central ${\rm Ir_2O_2}$ plaquette (open
    symbols). The inset shows the definition of $\phi$, where Ir and the O atoms
    are represented by filled blue and red circles respectively.
    The vertical dashed lines represent the approximate
    $\phi$ values for {\kiro} and bond {\it B1} of {\niro}, and the corresponding
    magnetic couplings are shown by filled symbols. The solid lines are
    guide to eye.
    }
   	\label{fig:gamma_theta}
\end{figure}

In order to understand the origin of such large $\Gamma$'s,
we study the (distinct) structural differences with {\niro}. The
characteristic features of the honeycomb planes in {\kiro} are
negligible trigonal
distortions (due to large inter-layer separations) and a regular
$120^{\circ}$ arrangement of O-O links centered at Ir-Ir bonds when
projected on the plane of Ir atoms (see
Fig. \ref{fig:str_kiro}(b)). 
Therefore, starting from {\kiro}, an approximate lower symmetry
structure similar to {\niro}
can be obtained in two steps: First, within the considered Ir$_2$O$_{10}$
unit, the four outer O-O links are rotated about an axis perpendicular
to the Ir plane and centered between the respective O-O links
such that the structural unit retains the $C_{\rm 2h}$
symmetry, albeit without the $C_3$ symmetry about the '$c$' axis at the Ir sites. 
Due to the $C_{\rm 2h}$ symmetry,
these octahedral distortions can be quantified by a single parameter:
twist angle $\phi$ which measures the deviation of the neighboring O-O links with
respect to the O-O link in the central Ir$_2$O$_2$ plaquette (see the inset of Fig. \ref{fig:gamma_theta}). Second,
introducing trigonal distortions with the trigonal axis perpendicular to
Ir plane.
We consider clusters with
different values of $\phi$ and trigonal distortion.
For consistency and
comparison, same point charge embedding was used for all the 
cases (see Methods).

The dependence of the NN magnetic couplings on $\phi$ is shown in Fig. \ref{fig:gamma_theta}. 
The strength of the Kitaev exchange ($\Gamma_{yz}$) increases
(decreases) away form the constrained case ($\phi = 0$) while
$\Gamma_{xy}$ changes sign. 
The NN Heisenberg exchange $J$, on the other hand, has a qualitatively
similar behavior as $K$, and lies in the range of 2 meV and 3.7 meV (not
shown). The corresponding values for the Ir$_2$O$_{10}$ structural units
of {\kiro} and {\niro} are also
shown. The strength of the $K$ and $J$ parameters are substantially larger
in the presence of trigonal distortion, as evidenced from the values for
{\niro} (filled symbols at $\phi \approx 8^{\circ}$ in
Fig. \ref{fig:gamma_theta}) and also noted earlier \cite{satoshi2016}.
We emphasize that the magnetic exchange couplings for the Ir$_2$O$_{10}$ units
in Fig. \ref{fig:gamma_theta} are obtained within the MRCI framework, and thus provide a
reliable qualitative estimate on their evolution with $\phi$.  However, a
full chemical treatment of the NN octahedral units is required to obtain
a good quantitative estimate of magnetic exchange interactions, as done
for the values in Table 1 (see Methods for details). 

Presence of smaller Na ion at the center of hexagons in the honeycomb
layer in {\niro} leads to 
deviations from the $120^{\circ}$ order. The O atoms reorganize
themselves to lower the total energy. This is confirmed by comparing
the total energy of {\niro} and an equivalent structure derived from
{\kiro} such that the unit cell volume, and Ir-Ir and Ir-O distances 
for these two structures are
comparable (see Methods). 
{\niro} is found to be approximately 18 meV/atom lower in
energy, thus establishing that the presence of larger 
intra-honeycomb-layer K ions leads to constraints in the
relative orientation of the O atoms which, in turn,
leads to large $\Gamma$'s.
Similar nearly-$120^{\circ}$ arrangement for the O-O links is also found
in the hyper-honeycomb compound $\beta$-${\rm Li_2IrO_6}$ which also 
has somewhat larger $\Gamma$'s \cite{Majumder18}.

These results suggest that the real space $C_3$ symmetry at the
Ir sites is intricately related to the strong magnetic anisotropy.
Even small deviations significantly influence the magnetic couplings, 
suggesting strong entanglement between real space and spin space.
As a result, the NN magnetic interactions are strongly frustrated, highly
anisotropic in spin space and dominated by the $\Gamma$-terms.
While a sizable  $\Gamma_{yz}$ in such a situation is
plausible for large trigonal distortions \cite{chaloupka2015}, it is
interesting that unusually
large $\Gamma_{ij}$'s in {\kiro} is realized even in the absence of sizable
trigonal distortions.

It is important to note the fully anisotropic nature of the above
    $J$-$K$-$\Gamma$ model. In comparison, some of the previously studied models
consider only one component of $\Gamma$ per bond
\cite{gamma_SL2017,Catuneanu2018,Lampen-Kelly2018,Gohlke2018}.
On the other hand, in the fully anisotropic model for 
 {\niro} \cite{yadav2018}
and {\arucl} \cite{Yadav16}, the $\Gamma$ terms turn
    out to be small. Nevertheless, they may have important consequences
    for magnetism \cite{Catuneanu2018,Lampen-Kelly2018,Gohlke2018}. 
In this regard, the implications of such large $\Gamma$'s,
especially on the stability of a SL state, is particularly interesting.

\subsection{Phase Diagram for the $K$-$J$-$\Gamma$ model}
\begin{figure}
    \centering
    \includegraphics[angle=0,width=0.49555\textwidth]{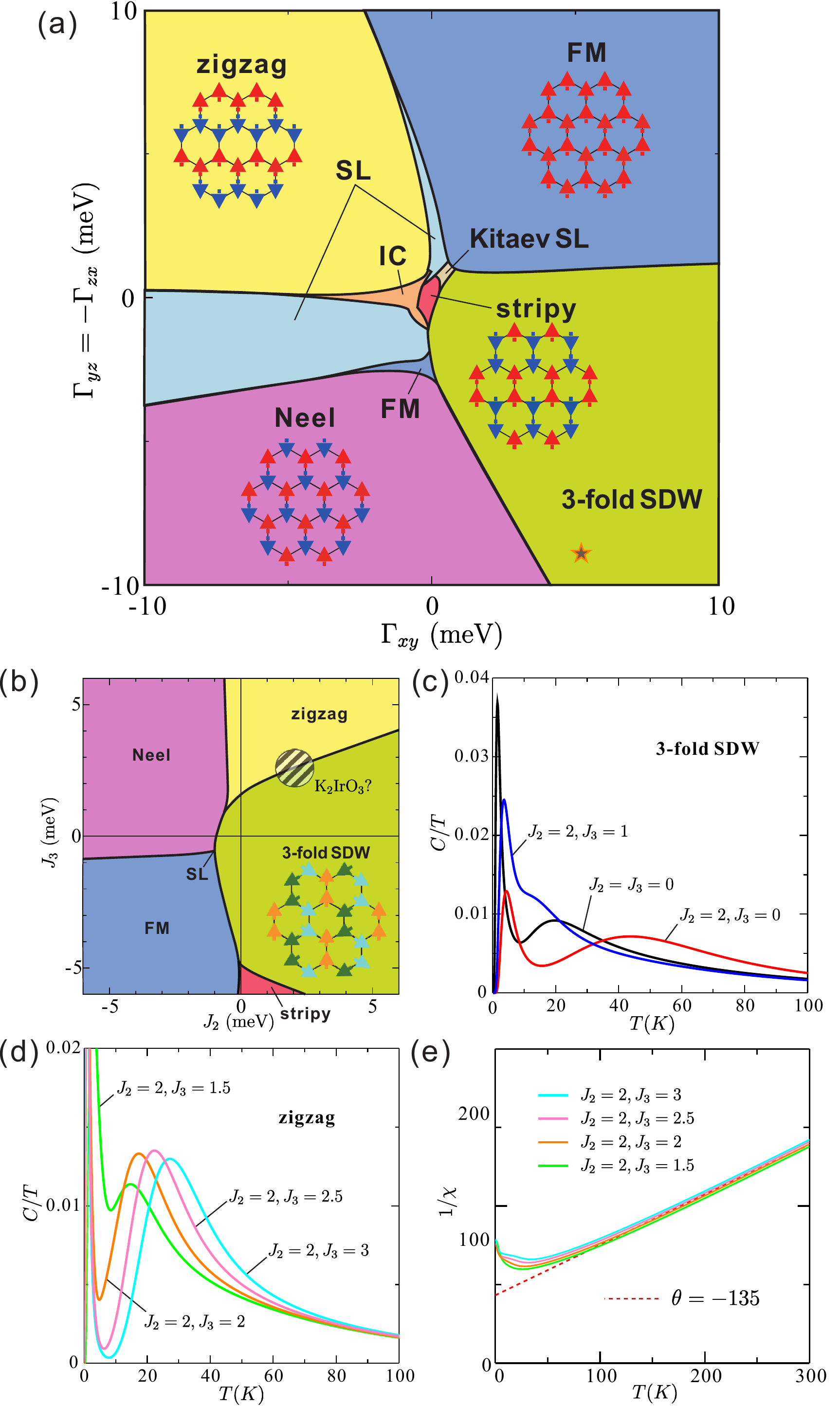}
    \caption{\small
    Ground state phase diagram by ED with a 24-site cluster:
    (a) in the $\Gamma_{xy} - \Gamma_{yz}$ plane using the MRCI values of
    $J$ and $K$; (b) in the $J_2$-$J_3$ plane using the MRCI values of
    $J$, $K$, $\Gamma_{xy}$, and $\Gamma_{yz}$. Schematic spin configurations
    are also shown. The star symbol in (a) indicates the position of MRCI parameter
    set and it corresponds to the origin in (b).
    A realistic range for {\kiro} is located with shaded area in (b).
    Specific heat for the (c) 3-fold SDW and (d) zigzag phases, obtained by
    the ED calculations with a 12-site cluster. 
    (e) Inverse magnetic susceptibility for the most likely realistic values of
    the extended range Heisenberg couplings $J_2$ and $J_3$. The dotted
    line is $\chi=2.1/(T-\theta)$ with $\theta=-135$ K.
        }
    \label{fig:ED}
\end{figure}

We employ exact diagonalization (ED) calculations for the lattice realization
of the fully anisotropic $K$-$J$-$\Gamma$ model described by Eq.
(\ref{eqn:bilinear_qc}) on a 24-site cluster \cite{chaloupka10,Katukuri14,Nishimoto16,Yadav16}.
The resulting magnetic phase diagram in the $\Gamma_{xy}
- \Gamma_{yz}$ plane is very rich due to
competing magnetic interactions, shown in Fig. \ref{fig:ED}(a)
where the MRCI values of $J=1.3$ and $K=-6.3$ were fixed. It includes
six ordered phases: ferromagnetic (FM), N\'eel, zigzag, stripy, 3-fold spin-density wave (SDW),
and incommensurate (IC) (see the Supplemental Material for details). 

The most remarkable feature is that a ($\Gamma$-driven) SL state is found to be stable
in a large region of the phase diagram although a stripy state is realized 
at small values of $\Gamma_{ij}$. It suggests an alternative route to realize
a SL state in spin-orbit driven Mott insulators on a honeycomb lattice.
Especially, it is striking that a Kitaev-type SL is recovered 
between the FM and stripy phases (see the Supplemental Material).
The FM$-$Kitaev-SL$-$stripy path also appears when $K/J$ is varied
in the $K$-$J$ model \cite{chaloupka10}, 
implying that the ratio $K/J$ could be effectively controlled by
the $\Gamma$ terms. The remaining SL region is characterized as
a frustration-induced ``conventional'' disordered state.
A relatively wide region of SL phase exists for $\Gamma_{ij} < 0$; whereas,
with increasing $\Gamma_{ij} >0$, a SL state is stable only for small
range, eventually leading to the FM order. This is
consistent with an earlier work for the $K$-$\Gamma$ model \cite{Catuneanu2018}.

Let us now consider the situation for {\kiro}. As indicated in Fig. \ref{fig:ED}(a),
the {\it pure} MRCI parameter set ($\Gamma_{xy}=5.2$ meV, $\Gamma_{yz}=-8.9$ meV)
stands on the 3-fold SDW ordered phase. However, extended range
Heisenberg couplings $J_2$ and $J_3$ are known to be important for
a honeycomb lattice spin-orbit driven Mott insulators \cite{Katukuri14}.
Thus, we obtain a phase diagram considering $J_2$ and $J_3$ couplings
in addition to the MRCI parameter set, which is shown in Fig. \ref{fig:ED}(b).
To estimate realistic values of $J_2$ and $J_3$ for {\kiro}, we
turn towards the recent experimental observations.\cite{kiro_expt18_mag}
Key features of the experimental specific heat are a broad maximum around $30$K
and finite $C/T$ down to $1.8 K$.  As shown in Fig. \ref{fig:ED}(c)(d), the broad
peak at $\sim 30$K can be numerically reproduced by setting $J_2 \sim 2$ and
$J_3 \sim 2$-$3$. Given these values, the system is just near the boundary between
the zigzag and 3-fold SDW phases (see Fig. \ref{fig:ED}(b)). Therefore, 
a strong competition between different orders is naturally expected, 
possibly explaining 
why no long-range order has been observed down to $\sim 2$K. 
Actually, the related N\'eel temperature is found to be $T_{\rm N} <2$K
in the numerical calculations of specific heat.
The corresponding inverse magnetic susceptibility for the zigzag state in an external
magnetic field of 2T perpendicular to the honeycomb planes leads to
$\theta \sim -135$K (Fig. \ref{fig:ED}(e)) 
which is somewhat smaller than the experimental value of
$\theta \sim -180$K. \cite{kiro_expt18_mag}
Nonetheless, the zigzag state is the most probable ground state.

The synthesized {\kiroxy}
samples are non-stoichiometric and possess significant disorder in the
form of vacancies in the K-layer and presence of Ir/K occupancy at
the centers of the hexagons in the honeycomb layers. As was recently
shown \cite{Yadav2018_PRL,Yadav2019_ChemSc}, disorder,
especially vacancies and position of the inter-layer cation, can
significantly influence the magnetic
couplings, and may possibly drive the system towards a SL state. However, a detailed study
of such effects require considerations beyond the scope of the present
work.

\section{Conclusions}
 In summary, we have investigated the magnetic interactions and the possibility of QSL states in a
structural model for honeycomb lattice Mott insulator featuring $C_3$ symmetry at the TM ion sites by
considering the recently proposed structural model for {\kiro}.
We find that the resulting magnetic anisotropy is dominated by unusually
large off-diagonal anisotropic exchange couplings. 
Such large $\Gamma$'s are related to the $C_3$ PG
symmetry at the Ir sites, 
leading to constraints on the relative arrangement of O atoms. 
Despite small values of $K$ and $J$ as compared to {\niro},
the sizable $K/J$ ratio motivates a fully anisotropic $K$-$J$-$\Gamma$
model for {\kiro} leading to strong magnetic frustration. 
Large quantum fluctuations suppress the magnetic ordering down to
$\sim 2 K$.

A fully anisotropic  $K$-$J$-$\Gamma$
Hamiltonian with large $\Gamma$ terms is a generic feature of honeycomb Mott insulators with $C_3$ PG
symmetry at the TM-ion sites and has a rich magnetic phase diagram.
We highlight that the relative sign of $\Gamma_{ij}$ is critical for the
stability of a SL state. The most remarkable aspect is the hitherto unexplored
possibility that a SL can be stabilized even for small values of $\Gamma
<0$ when $K$ is ferromagnetic and the ratio $|K/J|$ is relatively large
($\gtrsim 5$). At the same time, the recognized
FM$-$Kitaev-SL$-$stripy path of the $K$-$J$ model is also recovered at
finite $\Gamma_{ij}$. While the search for a SL state in spin-orbit driven Mott
insulators so far is focused on tuning the ratio $K/J$, our findings
suggest that tuning the off-diagonal anisotropic couplings in materials
with $K<0$ should also be promising.

\section{methods}
\subsection{Density Functional Calculations.}

Density Functional Theory (DFT) calculations were performed using the Perdew-Burke-Ernzerhof (PBE) implementation of the
generalized gradient approximation (GGA) as implemented in the FPLO code
\cite{Klaus99}, version 18.52 \cite{fplo_web}. A $k$-mesh with $12 \times 12
\times 12$ intervals in the full Brillouin zone was used for numerical integration along with
a linear tetrahedron method. For the electronic properties, the `Atomic
Limit' (AL) implementation of the GGA+{\it U} functional was used.
The spin-orbit effects were considered within the 4-spinor formalism. For simplicity, the quantization axis was
chosen to be [0 0 1].

The structure optimization was performed for the internal parameters
only, utilizing the space group $P6_322$ (\# 182) and including the scalar
relativistic corrections.
The experimental values of the
external parameters (lattice constants): $a = b= 5.282$ {\AA}, $c=
13.544$ {\AA} \cite{kiro_expt18} were used. The residual force is less than 1 meV/{\AA} on each atom.
The optimal atomic positions are presented in the Supplemental
    Material. 

\subsection{Quantum chemistry calculations.}

To determine the strength of NN magnetic couplings for the 
DFT-optimized structure, quantum chemistry calculations were performed on a material
model consisting of two NN edge shared octahedra unit Ir$_2$O$_{10}$.
Additionally, the four octahedra sharing edge with the central unit along with eighteen NN K ions
were also explicitly included in the calculations to
account for the finite charge distribution in the immediate
neighborhood. The remaining solid-state surroundings were modeled by arrays of
point charges such that they reproduce the ionic Madelung potential in the
cluster region. Energy-consistent relativistic pseudopotentials along
with quadruple-zeta basis functions were used for the
Ir ions \cite{Figgen09} of  the central unit while an all-electron
quintuple-zeta  basis sets were employed for the bridging
O ligands \cite{Dunning89}. The remaining O atoms in the
two-octahedra central region were described using an all-electron basis sets of triple-zeta
quality \cite{Dunning89}. Ir$^{4+}$ sites belonging to the octahedra
adjacent to the reference unit were described as closed-shell Pt$^{4+}$
$t_{2g}^6$ ions, using relativistic pseudopotentials and valence
triple-zeta basis functions \cite{Figgen09}. Ligands belonging to these adjacent
octahedra which are not shared with the central (reference) unit were
modeled with a minimal all-electron atomic-natural-orbital basis sets
\cite{Pierloot95}. All occupied shells
of NN K$^+$ sites were represented by using pseudopotentials and each of
the K 4s orbitals were described with a single basis function \cite{Fuentealba82}. All quantum chemistry computations were performed
using the quantum chemistry package {\sc molpro}.

The magnetic couplings in Table \ref{table:couplings}  were obtained by
mapping the \textit{ab initio} data, obtained in the multi-reference
configuration-interaction (MRCI) and including spin-orbit effects, as
mentioned in the main text, onto an effective
spin Hamiltonian (see Eq. (1) in the main text) following the scheme
detailed in Ref. \cite{Yadav16}.

In the first step, complete-active-space self-consistent-field
(CASSCF) calculations \cite{Helgaker2000} were carried out 
for an average of the lowest nine singlet and nine triplet states,
essentially of $t_{2g}^5-t_{2g}^5$ character. Since CASSCF
calculations also account for superexchange processes of 
$t_{2g}^6-t_{2g}^4$ type in addition to NN $t_{2g}^5-t_{2g}^5$ direct
exchange, CASSCF wavefunctions also consist of a finite weight
contribution from inter-site excitations of $t_{2g}^6-t_{2g}^4$ type.
Single and double excitations from the transition metal $d$ ($t_{2g}$) and
bridging-ligand $p$ valence-shells were accounted for in the subsequent
multireference configuration-interaction (MRCI) computations.
The low-lying nine singlet and nine triplet states were considered
in the spin-orbit treatment, in both CASSCF and MRCI calculations. 
In the next step, the {\it ab initio} quantum-chemistry
data were mapped onto an effective spin Hamiltonian (Eq. (1) in
the main text) which involves only the lowest four spin-orbit states,
associated with the different possible couplings of the two NN 1/2
pseudospins. The other 32 spin-orbit
levels arising from the $t_{2g}^5-t_{2g}^5$ configuration involve
$j_{\rm eff} \approx 3/2$ to $j_{\rm eff} \approx 1/2$ excitations and
lie at significantly higher
energy \cite{Katukuri14,Katukuri16}. 
The mapping was performed following the procedure described in Ref.
[\cite{Yadav16}].
The magnetic couplings shown in Table 1 in the main text are obtained by
following this procedure.

\subsection{Tests for the role of $\mathbf{C_3}$ point group symmetry.}

To further test our claim that $C_3$ point group symmetry at each Ir site
is crucial to the strength of off-diagonal exchange couplings,
we performed an additional set of quantum chemistry calculations using an edge shared octahedra unit [Ir$_2$O$_{10}$]$^{2-}$
as the central region. We begin with a [Ir$_2$O$_{10}$]$^{2-}$ structural unit
obtained from an idealized crystalline structure displaying $C_3$ point
group symmetry about an axis perpendicular to the honeycomb plane
('$c$'-axis) at each Ir site. The structural unit, therefore, has
$C_{2h}$ point group symmetry. All adjacent Ir and
K sites were modeled as identical point charges to make the whole system charge neutral.
Using this model, {\it ab initio} spin-orbit calculations at the MRCI
level were carried out to obtain effective coupling parameters. 

In the next step, we reduced the symmetry of the [Ir$_2$O$_{10}$]$^{2-}$
structural unit such that only $C_{2h}$ symmetry of the cluster is
retained while the $C_3$ point group symmetry at each Ir site is lost.
This was obtained by keeping the central plaquette unchanged (i.e. no
modifications to Ir-Ir or Ir-O bond lengths and Ir-O-Ir bond angles
within this plaquette). However, the 
other sets of O atoms (representing the other two plaquette at each
site) were rotated about an axis passing through the center of the O-O
links and perpendicular to the hexagonal plane spanned by the Ir atoms.
The positions of all such O atoms are connected by the two-fold
rotation about the Ir-Ir axis and the mirror plane perpendicular to the
Ir-Ir bond, which defines the $C_{2h}$ symmetry. The
resulting structures can therefore be characterized by a single ``twist angle"
$\phi$ (see Fig 2, main text). The structural positions of all the
atoms for $\phi = 0$ and $\phi = 10^{\circ}$ is presented in Table
S4. 
The magnetic couplings were then obtained for different structures
corresponding to different values of $\phi$ by keeping the point charge
embedding same as in the first step ($\phi = 0$).

The modification of the O sites corresponding to the NN plaquette affects the Ir $d$ orbitals in a way that results in 
an increase in Kitaev exchange while decreasing the off-diagonal exchange components at the same time.
MRCI spin-orbit results obtained for two site clusters with varying $\phi$ are shown in Fig. 2 in the main text.
The model structure with $\phi=0$ corresponds to the crystalline structure proposed for K$_2$IrO$_3$ while the model structure
with $\phi=10^{\circ}$ can be compared with the crystal structure of Na$_2$IrO$_3$.

Basis sets of the same quality as discussed in the previous section were
used for the structural unit in these calculations.
Additionally, to access the reliability of these test calculations we performed checks by changing the embedding to the one 
with a lower symmetry arising due to the distortions. We find that the qualitative trend
obtained in these calculations remains the same as shown in Fig. 2 in
the main text.

\subsection{Exact diagonalization calculations.}

To investigate the lattice magnetic structure, we calculated the static spin-structure factor
\begin{equation}
S({\bf Q})=\frac{1}{N}\sum_{ij} \langle {\tilde {\bf S}}_i\cdot {\tilde {\bf S}}_j \rangle \exp[i {\bf Q}\cdot ({\bf r}_i-{\bf r}_j)],
\end{equation}
where $N$ is the number of sites in a periodic cluster and ${\bf r}_i$
is the position of site $i$. The structure
factors $S({\bf Q})$ for representative momenta (first Brillouin zone)
in different phases is shown in the Supplemental Material. 
Distinction between different states realized
in the phase diagram (Fig. \ref{fig:ED}(a)), especially between the ordered and
incommensurate phase, and between the `conventional' (non-Kitaev-type)
and Kitaev-type spin liquid (SL) states is illustrated. Further, the
implication of finite cluster size is also discussed. 

The phase boundaries were obtained by checking the second derivatives of the
ground-state energy $E_0/N$ and the total spin $2S_{\rm tot}/N$ with
respect to particular parameters. The critical $\Gamma$'s values for the
phase transitions were
estimated from the peak positions in the second derivative of the
ground-state energy $-\partial^2 E_0/\partial \Gamma_{yz}$ (see
Supplemental Materials for a 
detailed discussion and illustration of the 
SL$-$IC$-$stripy$-$Kitaev SL$-$FM phase transitions. 

To characterize the topological nature of the spin liquid state, the
hexagonal plaquette operator was considered, which is defined as \cite{Kitaev06}:
\begin{equation}
    O_{\rm h}=2^6 {\tilde{S}}_1^x {\tilde{S}}_2^y {\tilde{S}}_3^z
    {\tilde{S}}_4^x {\tilde{S}}_5^y {\tilde{S}}_6^z \,,
\label{eqn:O_h}
\end{equation}
where the site labels (subscripts) correspond to the six sites in a hexagonal
ring and the link labels (superscripts) denote the corresponding bond anisotropy for
bonds away from the hexagonal ring (see Supplemental Materials for details).

{\it Acknowledgment.}
We thank Yogesh Singh for the structural data which
motivated this study, and Liviu Hozoi for helpful discussions.
Part of this work is supported by the DFG through SFB 1143 project A05.
RR and MR acknowledge financial
support from the European Union (ERDF) and the Free
State of Saxony via the ESF project 100231947 and 100339533 (Young
Investigators Group Computer Simulations for Materials
Design - CoSiMa.)
We also thank Ulrike Nitzsche for technical support.

\bibliography{kiro}

%
%
%
\renewcommand{\thetable}{S\arabic{table}}
\renewcommand{\thefigure}{S\arabic{figure}}
\setcounter{section}{0}
\setcounter{subsection}{0}
\setcounter{figure}{0}
\setcounter{table}{0}
\setcounter{equation}{0}

\include{SM}

\end{document}

%% file: SM.tex
\section*{Supplemental Material}
\section{Structural Details and electronic properties}
\subsection{Optimized structure}
The optimal atomic positions for {\kiro} are listed in Table \ref{table:str_rr}, and
are in good agreement with the reported values \cite{kiro_expt18}.
Note that the nearest neighbor octahedra are of slightly
different sizes as implied by different Ir-O bond lengths. Therefore, 
at each Ir-Ir bond, inversion and mirror symmetry are absent.
Consequently, the Ir$_2$O$_{10}$ clusters do not have $C_{2h}$
symmetry. However, deviation from the $C_{2h}$ symmetry is only
marginal as the Ir-O bond lengths differ by $\lesssim 0.19$\%.
Therefore, $C_{2h}$ symmetry is a reasonable approximation and is used
in the quantum chemistry calculations.


\begin{table}[ht!]
	\caption{Optimal atomic positions and related structural parameters
        in {\kiro}. The external parameters are kept fixed to the experimental
        value \cite{kiro_expt18}: space group $P6_322$ (\# 182) with $a =
        b = 5.282$ {\AA}, $c=13.544$ {\AA}. 
        }
    \label{table:str_rr}
    \begin{tabular*}{0.48\textwidth}{p{2.6cm} p{2.0cm} p{4.00cm} }
    \hline
    \hline
        {\bf Atom.} & {\bf Wyk-pos} & {\bf Coordinates} ($x/a,y/b,z/c$) \\
    \hline
    \hline

    Ir1      & $2b$ &	(0, 0, 1/4) \\
    Ir2      & $2c$ &	(1/3, 2/3, 1/4) \\
    K1       & $2c$ &	(2/3, 1/3, 1/4) \\
    K2       & $6g$ &	(0.3404, 0.3404, 1/2) \\
    O        & $2a$ &	(0.3328, 0.0593, -0.3442) \\
    
    \hline
    \hline
       {\bf Param.} &    & {\bf Values} \\
    \hline
    \hline
    $d_{\rm Ir - Ir}$ ({\AA})  &      &   3.0496 \\
    $d_{\rm Ir - O}$ ({\AA})  &      &   2.065, 2.069 \\
    $d_{\rm Ir-layers}$ ({\AA})      &   &   6.77 \\
    $\angle {\rm Ir-Ir-Ir}$   &      &   120$^{\circ}$ \\
    $\angle {\rm Ir-O-Ir}$    &      &   95.06$^{\circ}$ \\
    \hline
    \end{tabular*}
\end{table}

\subsection{Electronic Properties}

\begin{figure}[h]
\centering
\includegraphics[width=1.02\linewidth]{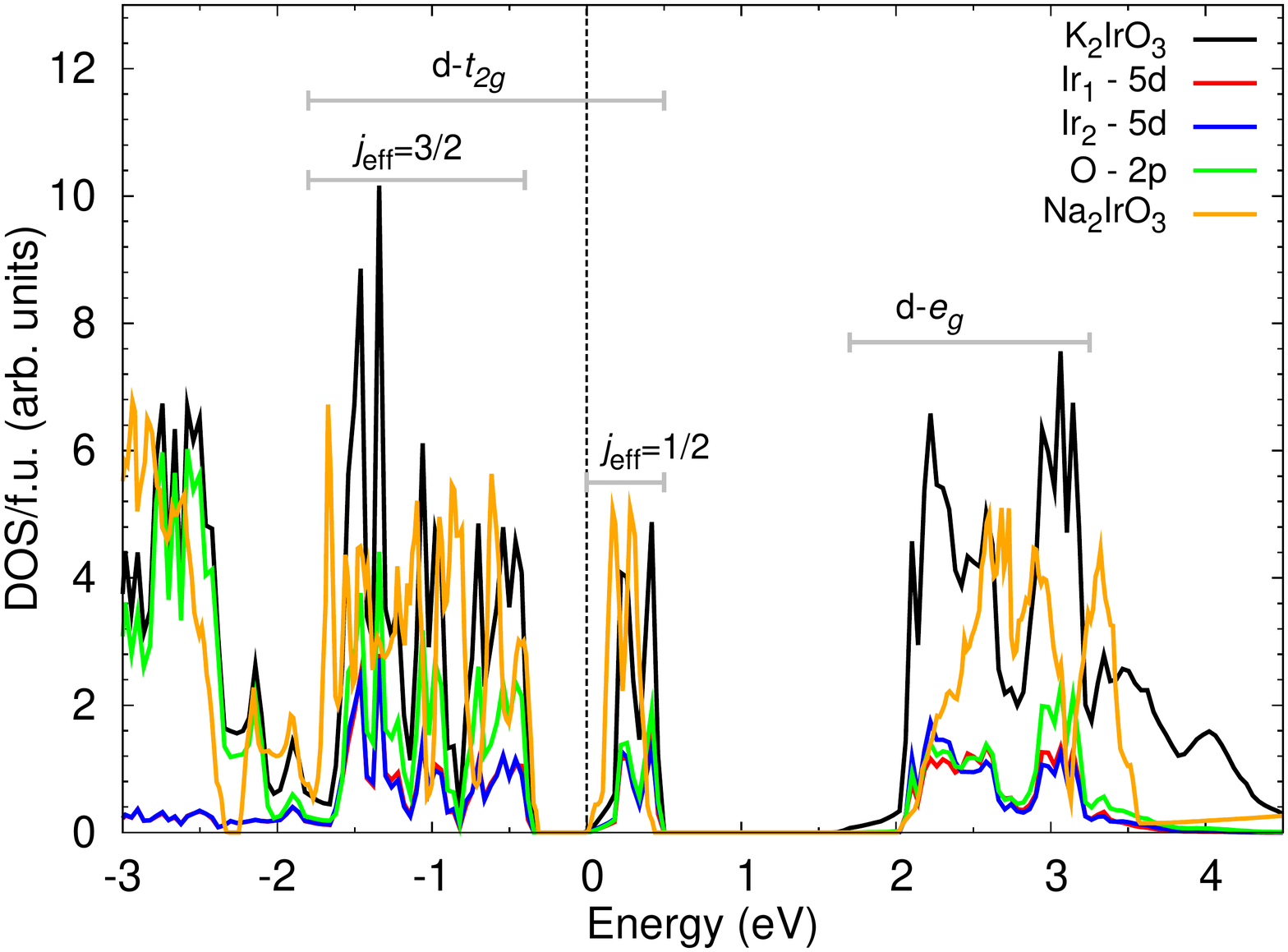}
\caption{
    Total (spin-up + spin-down) density of states (DOS) per formula unit
    (f.u.) for the spin-orbit driven Mott Insulator
    {\kiro} along with the contributions of Ir-$5d$ and O-$2p$
    states as obtained within the GGA+{\it U} scheme and considering
    spin-orbit effects, with $U_{{\rm Ir}-5d} = 1.2$ eV and  $J_{{\rm
    Ir}-5d} = 0.3$ eV. For comparison, the total DOS for {\niro}
    with the same $U$ and $J$ values is also shown. The Fermi energy is
    shown by dashed vertical line. 
}
\label{fig:pdos}
\end{figure}

Figure \ref{fig:pdos} shows the total and partial density of
states (DOS) for {\kiro} with  $U = 1.2$ eV and  $J= 0.3$ eV applied
to the Ir-$5d$ states. For comparison, the total DOS for
{\niro} with the reported lattice parameters \cite{choi12} and
optimized internal parameters, and with the same $U$ and $J$ values for the Ir-$5d$ states is
also shown. The $d$-${t_{2g}}$ and $d$-${e_g}$ states are also marked. The
clear splitting of the $d$-${t_{2g}}$ bands into $j_{\rm eff} =3/2$ and
$1/2$ states is evident, implying that {\kiro} is a spin-orbit
driven Mott insulator. The ground state properties, especially the
magnetic moments and residual charges, are comparable with {\niro}, as
shown in Table \ref{table:el_prop_compar}.
        
\begin{table}[hb!]
    \caption{Comparison of the ground state magnetic properties of
    {\kiro} and {\niro} obtained within the GGA +{\it U} scheme with
    spin-orbit coupling. $U=1.2$ eV and $J=0.3$ eV was applied to
    the Ir-$5d$ states.
    }
    \label{table:el_prop_compar}
    	\begin{tabular*}{0.49\textwidth}{p{3.2cm} p{3.2cm} p{2.60cm} }
       	\hline
       	\hline
            {\bf Param.}  & {\bf {\kiro}} & {\bf {\niro}} \\
       	\hline
       	\hline
            $m_S^{\rm Ir}$ ($\mu_{\rm B}$)   & 0.574, 0.569  &	0.431 \\
            $m_S^{\rm O}$ ($\mu_{\rm B}$)    & 0.095  & 0.062, 0.061 \\
            $m_S^{\rm K/Na}$ ($\mu_{\rm B}$) & 0.005, 0.002  & 0.003, 0.001 \\
            $m_S^{\rm tot}$ (per f.u., $\mu_{\rm B}$) & 0.864  & 0.620 \\
            $m_L^{\rm Ir}$ ($\mu_{\rm B}$)  & 0.862, 0.855  &	0.584 \\
        \hline
       	\hline
       	\end{tabular*}
\end{table}

\subsection{Comparison with an equivalent ${\rm Na_2IrO_3}$ structure
--- role of K-ions}

To ascertain the relation between the $C_3$ point group symmetry at the
Ir sites and the presence of the K ions at the center of the hexagons in
the honeycomb layer, we
performed DFT calculations with the following assertion: if the $C_3$ point group
symmetry is a direct consequence of the larger K$^+$ ions, presence of smaller non-magnetic cation, such as Na$^+$,
would allow the O atoms to reorganize themselves to a lower symmetry
structure and, subsequently, lower the total
energy. Indeed, Na$_2$IrO$_3$ crystallizes in the space group
$C2/m$ \cite{choi12}, which does not have the 
relative $120^{\circ}$ arrangement of the O-O links (shown in Fig. 1(b) in the main text).

The truthfulness of this assertion was checked by comparing the ground
state energies (within GGA) of the known structure of {\niro} \cite{choi12}, but with optimized atomic
positions, with an equivalent structure derived from {\kiro} with the
$C_3$ point group symmetry at the Ir sites, labeled as {\niro$^{*}$}.
The equivalency between the two structures correspond to comparable Ir-Ir
and Ir-O distances and the unit cell volume.
{\niro$^{*}$} was obtained from {\kiro} in the following steps:
\begin{enumerate}
    \item  Replace all K ions by Na ions.
    \item Tune $a$ (and $b$) such that the Ir-Ir distances are comparable to that of
    {\niro}. 
    \item  Reduce $c$ such that the unit cell volumes match while the
    inter-layer separations and the Ir-O distances are comparable to {\niro}. 
    \item Optimize the atomic positions of the resulting structure. 
\end{enumerate}
The atomic positions and structural details of both the
structures are presented in Table \ref{table:energy_compar}.

\begin{table}[b]
	\caption{Comparison of the structural parameters between {\niro
    $^*$} and {\niro}. `$*$' denotes that this structure was obtained from
    {\kiro} such that the Ir-Ir distances and the unit cell volume are comparable with
    {\niro}.
    }
    \label{table:energy_compar}
    	\begin{tabular*}{0.490\textwidth}{p{2.15cm} p{3.4cm} p{2.90cm} }
       	\hline
       	\hline
            {\bf Atom.}  & {\bf {\niro}$^*$} & {\bf {\niro}} \\
       	\hline
       	\hline

        Ir1     &(0, 0, 1/4) &	(0, -0.3338, 0) \\
        Ir2     &(1/3, 2/3, 1/4) &	- \\
        Na1     &(2/3, 1/3, 1/4) &	(0, 0, 0) \\
        Na2     &(0.3414, 0.3414, 1/2) &	(0, -0.1598, 1/2) \\
        Na3     &-               &	(0.3404, 0.3404, 1/2) \\
        O       &(0.3332, 0.0291, 0.6449) &	(-0.25284, 0.18314,\newline -0.21001) \\
        
        \hline
       	\hline
           {\bf Param.}     &  & \\
       	\hline
       	\hline
        Space Group                     & $P6_322$ (\# 182) &   $C2/m$ (\# 12) \\
        Volume/f.u. ({\AA}$^3$)         & 67.646            &   67.646 \\
        Energy difference/f.u. (meV)     & 107.89            &   0.0 \\
        $d_{\rm Ir - Ir}$ ({\AA})       & 3.12              &   3.12, 3.14 \\
        $d_{\rm Ir - O}$ ({\AA})        & 2.061, 2.0623      &   2.054, 2.061\\
        $d_{\rm Ir-layers}$ ({\AA})     & 5.338             &   5.614 \\
        $\angle {\rm Ir-Ir-Ir}$         & 120$^{\circ}$     &   120.12$^{\circ}$,119.76$^{\circ}$ \\
        $\angle {\rm Ir-O-Ir}$          & 98.48$^{\circ}$             &   99.344$^{\circ}$ \\
       	\hline
       	\end{tabular*}
\end{table}

    \subsection{Structures away from $C_3$ point group symmetry}
In order to test the effects of $C_3$ point group symmetry
on the magnetic couplings, quantum chemical calculations
were carried for [Ir$_2$O$_{10}$]$^{2-}$ structural units
with different values of the twist angle, $\phi$ (see
Methods for details). Away from $\phi=0$, the structural
units possess $C_{2h}$ but not the $C_3$ point group symmetry. Table \ref{table:str_phi} lists the atomic positions for the
[Ir$_2$O$_{10}$]$^{2-}$ structural units at $\phi=0$ and
$\phi=10^{\circ}$. The corresponding values of the magnetic couplings
were shown in Fig. 2.

\begin{table}[ht!]
    \caption{Atomic positions of the atoms in the ${\rm Ir_2O_{10}}$
    clusters for different values of the twist angle $\phi$. 
        }
    \label{table:str_phi}
    \begin{tabular*}{0.459\textwidth}{p{1.7cm} p{2.2cm} p{2.2cm} p{2.20cm} }
    \hline
    \hline
        {\bf Atom.} & $x$({\AA}) & $y$ ({\AA})  & $z$ ({\AA}) \\
    \hline
    \hline
        ${\mathbf \phi = 0} $ \\
    \hline
    \hline
    Ir1  &  0.000000  &   0.000000   &   0.000000   \\
    Ir2  &  3.049560  &   0.000000   &   0.000000   \\
    O1   &  1.524780  &   0.759692  &   1.170942   \\
    O2   &  1.524780  &  -0.759692  &  -1.170942   \\
    O3   & -1.420303  &   0.940652  &   1.170942   \\
    O4   & -0.104478 &   1.700344   &  -1.170942   \\
    O5   & -0.104477 &  -1.700344   &   1.170942   \\
    O6   & -1.420303  &  -0.940652  &  -1.170942   \\
    O7   &  4.469863  &   0.940652  &   1.170942   \\
    O8   &  3.154038  &   1.700344   &  -1.170942   \\
    O9   &  3.154037  &  -1.700345   &   1.170942   \\
    O10  &  4.469863  &  -0.940652  &  -1.170942   \\
    
    \hline
    \hline
        ${\mathbf \phi = 10^{\circ}} $ \\
    \hline
    \hline
     Ir1 &     0.000000 &  0.000000 &  0.000000  \\
     Ir2 &     3.049560 &  0.000000 &  0.000000  \\
     O1  &     1.524780 &  0.759692 &  1.170942  \\
     O2  &     1.524780 & -0.759692 & -1.170942  \\
     O3  &    -1.476267 &  1.060668 &  1.170942  \\
     O4  &    -0.048513 &  1.580328 & -1.170942  \\
     O5  &    -0.048513 & -1.580328 &  1.170942  \\
     O6  &    -1.476267 & -1.060668 & -1.170942  \\
     O7  &     4.525827 &  1.060668 &  1.170942  \\
     O8  &     3.098073 &  1.580328 & -1.170942  \\
     O9  &     3.098073 & -1.580328 &  1.170942  \\
     O10 &     4.525827 & -1.060668 & -1.170942  \\
    \hline
    \end{tabular*}
\end{table}

\section{Exact diagonalization calculations}

\subsection{Static spin-structure factor}

\begin{figure}[h]
\centering
\includegraphics[width=0.95\linewidth]{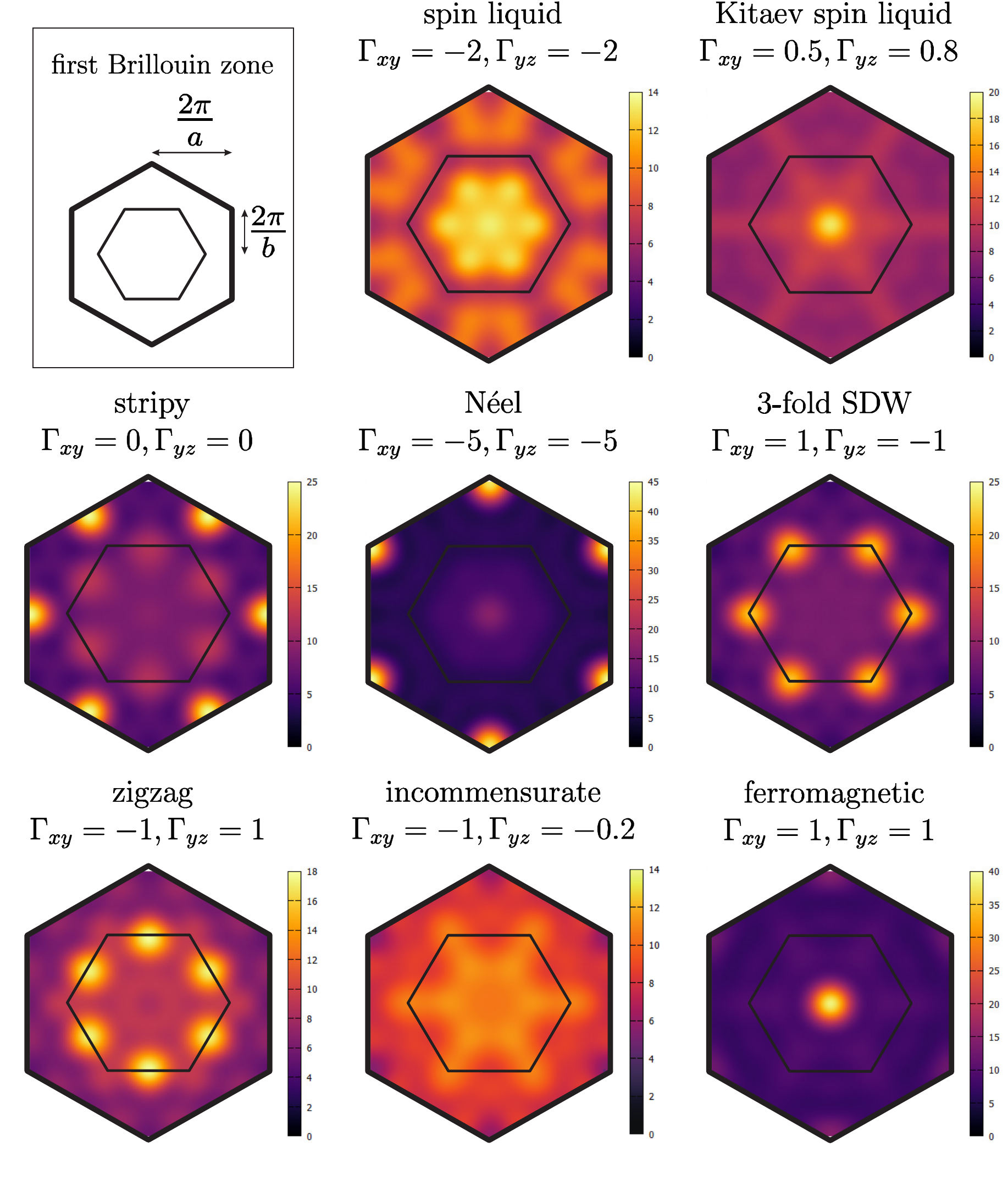}
\caption{
Spin-structure factor $S({\bf Q})$ for representative momenta in different phases.
}
\label{fig_SQ}
\end{figure}

The static spin-structure factor is defined as:
\begin{equation}
S({\bf Q})=\frac{1}{N}\sum_{ij} \langle {\tilde {\bf S}}_i\cdot {\tilde {\bf S}}_j \rangle \exp[i {\bf Q}\cdot ({\bf r}_i-{\bf r}_j)],
\end{equation}
where $N$ is the number of sites in a periodic cluster and ${\bf r}_i$ is the position of site $i$. In Fig.~\ref{fig_SQ} we show the structure
factors $S({\bf Q})$ for representative momenta (first Brillouin zone)
in different phases. Typical $\Gamma_{xy}$ and $\Gamma_{yz}$ values were
chosen for each phase. For the ordered phases the magnetic structure can
be determined from the reciprocal-space Bragg-peak positions. In the
incommensurate (IC) phase the Bragg-peak positions are shifted with
varying $\Gamma_{xy}$ and $\Gamma_{yz}$ if the system is large enough.
However, our system size is $24$ and only discrete momenta are allowed.
In such the case, usually, the dominant IC Bragg-peak position is moved
from a IC momentum to another by transferring the weight with varying
$\Gamma_{xy}$ and $\Gamma_{yz}$. Also, it is worth noting the difference
between the structure factors of the `conventional' (non-Kitaev-type)
and Kitaev-type spin liquid (SL) states. In the `conventional' SL phase
$S({\bf Q})$ is structureless and the weight is widely distributed over
the reciprocal space, reflecting a disordered spin state. Whereas, in
the Kitaev-type SL phase the weight is very small for the whole ${\bf
Q}$ range because the spin-spin correlations except for the
nearest-neighbor bond are very small. A peak at ${\bf Q}={\bf 0}$ is due
to the finite-size effect. Further discussion about the spin-spin
correlations is given below.

\subsection{Quantum phase transition}

\begin{figure}[h]
\centering
\includegraphics[width=0.96\linewidth]{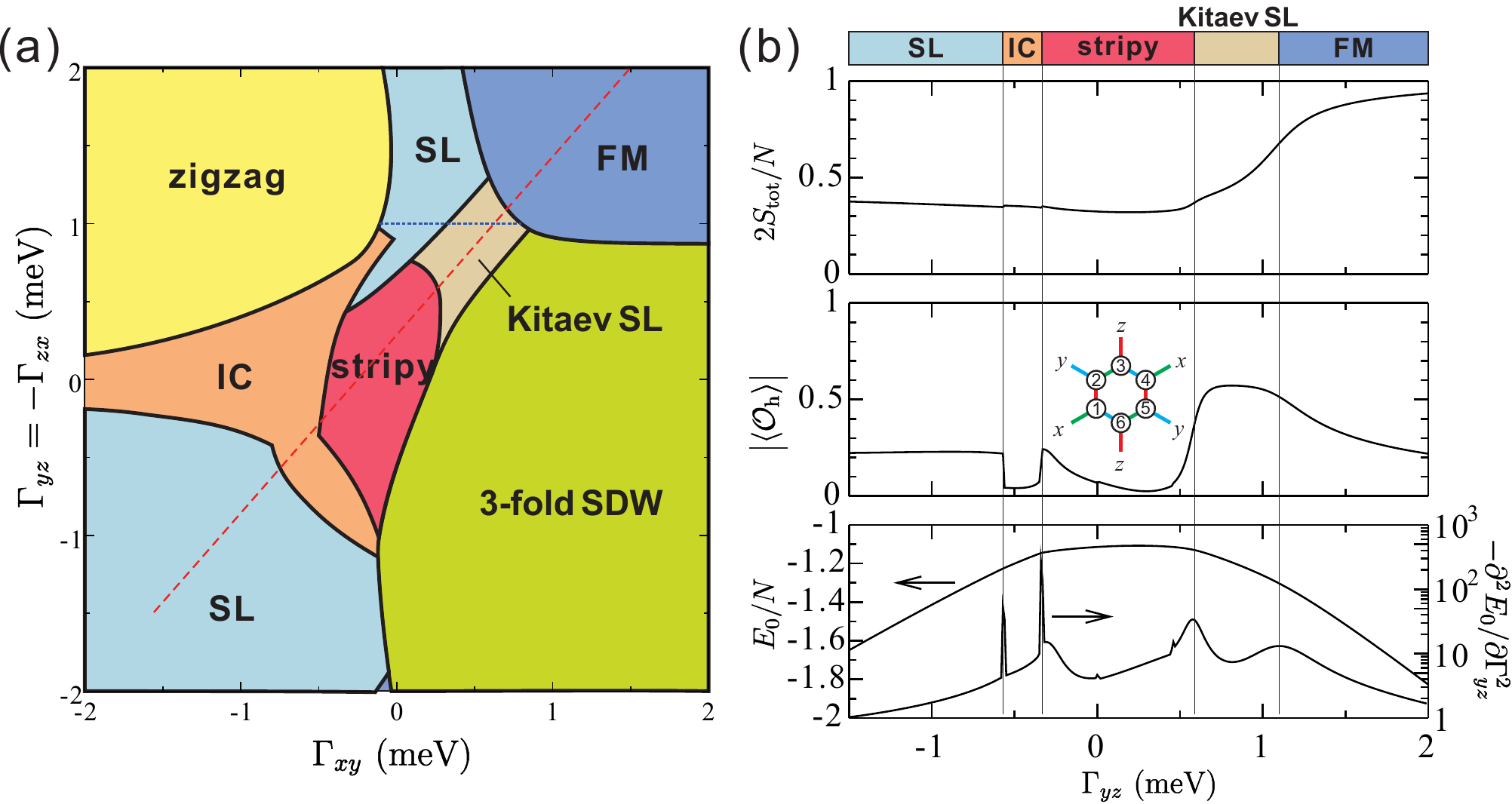}
\caption{
(a) Ground-state phase diagram around $\Gamma_{xy}=\Gamma_{yz}=0$,
extracted from Fig.~3 in the main text. (b) Representative physical
quantities for the 24-site periodic cluster as a function of
$\Gamma_{yz}$ along red dashed line
($\Gamma_{yz}=\frac{8}{7}\Gamma_{xy}+\frac{2}{7}$) in (a). Top: Total
spin, Middle: Expectation value of the hexagonal plaquette operator,
Bottom: Ground-state energy per site $E_0/N$ and its second derivative
$-\partial^2E_0/\partial \Gamma_{yz}^2$.
}
\label{fig_phase_transition}
\end{figure}

To find the phase boundaries, we checked the second derivatives of the
ground-state energy $E_0/N$ and the total spin $2S_{\rm tot}/N$ with
respect to particular parameters. In Fig.~\ref{fig_phase_transition}(b)
we illustrate four phase transitions (SL$-$IC$-$stripy$-$Kitaev SL$-$FM)
with varying $\Gamma_{xy}$ and $\Gamma_{yz}$ along dashed line in
Fig.~\ref{fig_phase_transition}(a). The critical $\Gamma$'s values were
estimated from the peak positions in the second derivative of the
ground-state energy $-\partial^2 E_0/\partial \Gamma_{yz}$; the
SL$-$IC$-$stripy$-$Kitaev SL$-$FM phase transitions occur at
$(\Gamma_{xy},\Gamma_{yz})=(-0.74875, -0.57)$, $(-0.54750, -0.34)$,
$(0.25750, 0.58)$, and $(0.71250, 1.10)$, respectively. The former two
transitions are of the first order, and the latter two are of the second
order or continuous. Furthermore, we considered the hexagonal plaquette
operator to check the topological property of the spin liquid state. The
plaquette operator is an indicative quantity for the Kitaev-type spin
liquid. It is defined as \cite{Kitaev06}:
\begin{equation}
O_{\rm h}=2^6 \tilde{S}_1^x \tilde{S}_2^y \tilde{S}_3^z \tilde{S}_4^x \tilde{S}_5^y \tilde{S}_6^z \,,
\label{O_h}
\end{equation}
where the labeling of links and sites is denoted in the middle panel of
Fig.~\ref{fig_phase_transition}(b). In the pure Kitaev limit ($K\gg
J,\Gamma$) the operator (Eq. \ref{O_h}) commutes with the Hamiltonian
and the expectation value of $\langle O_{\rm h} \rangle$ is exactly $\pm
1$. It is also known that it goes rapidly down to $\langle O_{\rm h}
\rangle \sim 0$ away from the Kitaev spin liquid regions
\cite{Yadav16}. As shown in Fig.~\ref{fig_phase_transition}(b), the
plaquette operator is significantly enhanced in a spin liquid region at
$0.25750<\Gamma_{xy}<0.71250$, $0.58<\Gamma_{yz}<1.10$. Thus, we
conclude that this spin liquid is of the Kitaev-type. This is also
consistent with almost weightless $S({\bf Q})$ shown above.

\subsection{Difference between Kitaev and non-Kitaev spin liquids}

\begin{figure}[htb]
\centering
\includegraphics[width=0.95\linewidth]{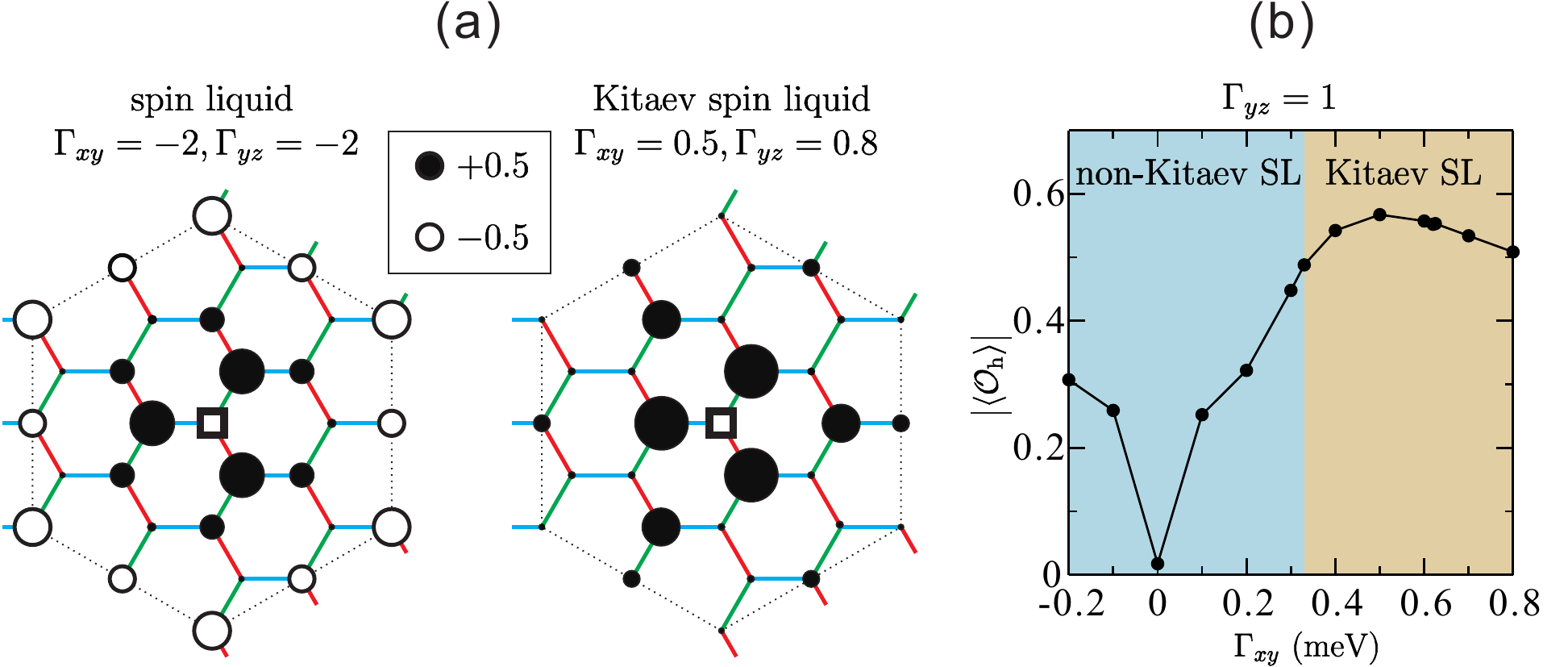}
\caption{
Spin-spin correlation function $\langle {\tilde {\bf S}}_i \cdot {\tilde {\bf S}}_j \rangle$ for non-Kitaev and Kitaev SL states 
in our phase diagram [Fig.~\ref{fig_phase_transition}(a)]. The reference site $i$ is denoted by open square and the value of 
$\langle {\tilde {\bf S}}_i \cdot {\tilde {\bf S}}_j \rangle$ at site $j$ is expressed by a circle. The filled and open circles mean 
positive and negative values of $\langle {\tilde {\bf S}}_i \cdot {\tilde {\bf S}}_j \rangle$, respectively. The diameter of each 
circle is proportional to the magnitude, i.e., $|\langle {\tilde {\bf S}}_i \cdot {\tilde {\bf S}}_j \rangle|$. 
(b) Expectation value of the hexagonal plaquette operator $|\langle O_{\rm h} \rangle|$ as a function of $\Gamma_{xy}$ 
at $\Gamma_{yz}=1$ fixed (along blue dotted line in
Fig.~\ref{fig_phase_transition}(a)).
}
\label{fig_spinspin}
\end{figure}

Sometimes, it is not easy to distinguish between Kitaev and non-Kitaev SL states because the phase transition 
is rather crossover-like. Nevertheless, they may be identified by looking at their spin-spin correlations (spin structure factor) 
and expectation value of the plaquette operator $\langle O_{\rm h} \rangle$.

In general, the Kitaev SL state is characterized by a rapid decay of the spin-spin correlations: in the Kitaev limit, 
only the NN correlations are finite and longer-range ones are zero; accordingly, as shown in Fig.~\ref{fig_SQ}, 
the static spin structure factor has a single ${\bf Q}={\bf 0}$ peak, the weight of which is much smaller than that for 
a FM state, in a finite-size cluster. On the other hand, the spin-spin correlations for a non-Kitaev SL are also not 
long ranged but the decay length are typically much larger, i.e., like a power-law decay, than that for the Kitaev SL; 
thus, a structureless $S({\bf Q})$ is obtained. This can be simply confirmed by considering the spin-spin correlations 
in the real space. In Fig.~\ref{fig_spinspin}(a) the real-space spin-spin correlations for non-Kitaev and Kitaev SL states 
are compared. We can obviously see that the correlation for Kitaev SL decays very rapidly and it is very small for 
longer distances than one lattice spacing; that for non-Kitaev SL decays much more slowly with distance.

We also show expectation value of the hexagonal plaquette operator $|\langle O_{\rm h} \rangle|$ around the phase 
boundary between non-Kitaev and Kitaev SL phases in Fig.~\ref{fig_spinspin}(b). A steep increase of $|\langle O_{\rm h} \rangle|$ 
from non-Kitaev towards Kitaev SL phases is clearly seen.

\subsection{Spin structure of the 3-fold SDW state}

\begin{figure}[htb]
\centering
\includegraphics[width=0.95\linewidth]{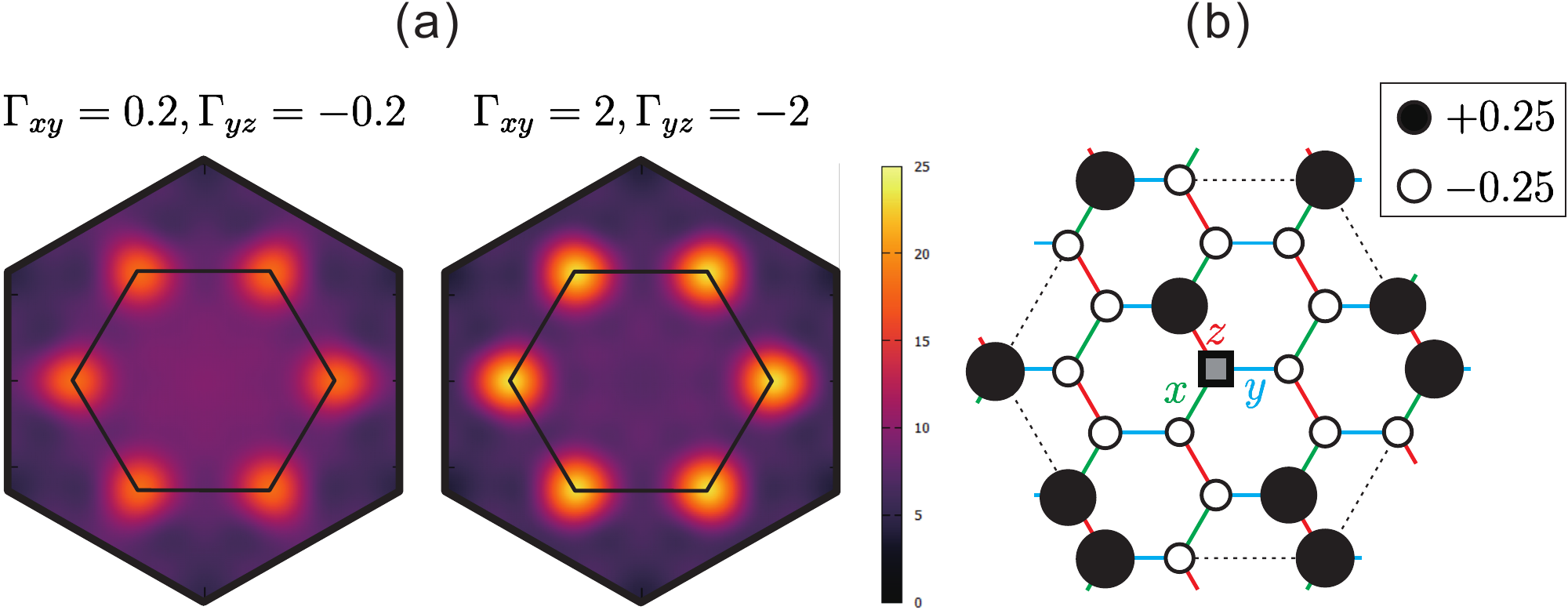}
\caption{
(a) Spin-structure factor $S({\bf Q})$ for $\Gamma_{xy}=0.2$, $\Gamma_{yz}=-0.2$ and $\Gamma_{xy}=2$, $\Gamma_{yz}=-2$, 
where the system is in the 3-fold SDW state. (b) Real-space spin-spin correlation function 
$\langle {\tilde {\bf S}}_i \cdot {\tilde {\bf S}}_j \rangle$  for a 3-fold SDW state ($\Gamma_{xy}=5$, $\Gamma_{yz}=-5$). 
The reference site $i$ is denoted by open square and the value of 
$\langle {\tilde {\bf S}}_i \cdot {\tilde {\bf S}}_j \rangle$ at site $j$ is expressed by a circle. The filled and open circles mean 
positive and negative values of $\langle {\tilde {\bf S}}_i \cdot {\tilde {\bf S}}_j \rangle$, respectively. The diameter of each 
circle is proportional to the magnitude, i.e., $|\langle {\tilde {\bf S}}_i \cdot {\tilde {\bf S}}_j \rangle|$. To detect the 
symmetry-broken state, the reference site is pinned by an infinite local magnetic field.
}
\label{fig_3foldSDW}
\end{figure}

As shown in our phase diagram [Fig.~3(a) in the main text], the ground state for a wide range of $\Gamma_{xy}>$ and 
$\Gamma_{yz}<$ is characterized as a 3-fold SDW phase. The spin-structure factor $S({\bf Q})$ for $\Gamma_{xy}=0.2$, 
$\Gamma_{yz}=-0.2$ and $\Gamma_{xy}=2$, $\Gamma_{yz}=-2$ are shown in Fig.~\ref{fig_3foldSDW}(a). The set 
$\Gamma_{xy}=0.2$, $\Gamma_{yz}=-0.2$ belongs to the 3-fold SDW ground state but is in the vicinity to neighboring 
stripy phase. Therefore, the peak structure in $S({\bf Q})$ is still somewhat blurred. Roughly speaking, this phase is 
more stabilized as $\Gamma_{xy}\approx-\Gamma_{yz}$ increases. We have also confirmed that the 3-fold SDW state 
is maintained up to the limit of $\Gamma_{xy}\approx-\Gamma_{yz}=\infty$.

In order to see the spin structure of 3-fold SDW state, we calculate the real-space spin-spin correlation function 
$\langle {\tilde {\bf S}}_i \cdot {\tilde {\bf S}}_j \rangle$. The result is plotted in Fig.~\ref{fig_3foldSDW}(b). A pinning 
is achieved by applying a infinite-strength magnetic field along $z$-axis on a site (referred as a reference site in 
Fig.~\ref{fig_3foldSDW}(b)) to directly detect a symmetry-broken state with a periodic cluster. We can clearly see a 
SDW structure with 3-fold oscillation perpendicular to the $z$-bond.

\flushbottom
\newpage
